\documentclass[%
reprint, onecolumn, superscriptaddress,
showpacs,preprintnumbers,
nofootinbib,
 amsmath,amssymb,
 aps,
pra,
]{revtex4}
\usepackage{mathrsfs}
\usepackage{graphicx}
\usepackage{epstopdf}
\usepackage{dcolumn}
\usepackage{bm}
\usepackage{color} 
\usepackage{CJK}

\def\la{{\langle}}
\def\ra{{\rangle}}

\newcommand{\beq}{\begin{equation}}
\newcommand{\eeq}{\end{equation}}
\newcommand{\beqa}{\begin{eqnarray}}
\newcommand{\eeqa}{\end{eqnarray}}

\newcommand{\cE}{{\cal{E}}}

\begin{document}

\title{Noise sensitivities for an atom shuttled by a moving  optical lattice via shortcuts to adiabaticity}

\author{Xiao-Jing Lu}
\affiliation{School of Electric and Mechatronics Engineering, Xuchang University, Xuchang
461000, China}
\author{Andreas Ruschhaupt}
\affiliation{Department of Physics, University College Cork, Cork, Ireland}
\author{Sof\'\i a Mart\'\i nez-Garaot}
\affiliation{Departamento de Qu\'\i mica F\'\i sica, UPV/EHU, Apdo 644 Bilbao 48080, Spain}
\author{J. Gonzalo Muga}
\affiliation{Departamento de Qu\'\i mica F\'\i sica, UPV/EHU, Apdo 644 Bilbao 48080, Spain}

\begin{abstract}
We find the noise sensitivities (i.e., the quadratic terms of the energy with respect to
the perturbation of the noise) of a particle shuttled by an optical lattice that moves according to  a shortcut-to-adiabaticity
transport protocol.  Noises affecting
different optical lattice parameters, trap depth,  position, and  lattice periodicity,  are considered. We find generic
expressions of the sensitivities for arbitrary noise spectra but focus  on the white-noise limit as a basic reference,
and on Ornstein-Uhlenbeck noise to account for
the  effect of  non-zero correlation times.
\end{abstract}
\maketitle
%
%
%
%

\section{Introduction}
The current  technical capabilities to control the translational motion of optical-lattice potential traps for atoms
make possible a plethora of  applications in quantum science and technology. We shall focus on the use of the lattice as a conveyor belt to transport atoms, although lattices may as well be moved for other purposes, e.g.
to study the stability of superfluidity \cite{Mun2007}   or, by  periodic driving (shaking), to control different aspects of single atoms  or many-body systems
\cite{Kiely2018,Eckardt2017}.
Optical traps are interesting to transport atoms because of several useful properties: the possibility to have hundreds or thousands of minima
(even more within hollow fibers \cite{Okaba2014,Langbecker2018}),
trapping forces that are much larger than in single beam optical tweezers, parameter flexibility including
time-dependent control, or the possibility to implement lattices that depend on the internal state \cite{Mandel2003}.
The atoms may be transported between a preparation area to a ``science chamber''
\cite{Middelmann2012,Okaba2014,Dinardo2016}, and  coherent control of individual atoms  has been demonstrated
towards on-demand positioning and delivery and the design of quantum registers \cite{Schrader2001, Kuhr2001,Kuhr2003,Miroshnychenko2003,Dotsenko2005,Miroshnychenko2006}.
Other applications include
guided  interferometry and precision measurement \cite{Lee2007,Steffen2012,Okaba2014,Langbecker2018}, quantum computation schemes via messenger atoms among distant register  qubits \cite{Calarco2004},  quantum random walks \cite{Dur2002,Alberti2014}, quantum simulators \cite{Jane2003},  catapulting (launching) atoms with specified velocities \cite{Kuhr2001,Schmid2006NJP},  creation of entangled states
\cite{Jaksch1999,Treutlein2006}, integrating  cold atoms with photonic platforms \cite{Kim2019}, or
implementing two-qubit quantum gates and gate arrays \cite{Jaksch1999,Brennen1999,Bloch2005}.

In most of the above applications fast transport processes are of interest,  e.g. to achieve high computational speeds,  to allow for many repetitions and
improve signal-to-noise ratios, or  to avoid decoherence,
but only as long as high fidelities with respect to desired final states are achieved.
Shortcuts to adiabaticity (STA) are a set of techniques devised to speed up slow adiabatic operations that help to design fast and robust operations, see
\cite{Torrontegui2013,Guery2019} for  review. In particular, STA have been applied to design fast transport operations that leave the
final state unexcited \cite{Couvert2008,Schmiedl2009,Masuda2010,Torrontegui2011}, or atom launching and stopping \cite{Torrontegui2011,Tobalina2017}, see further references for abundant  work on STA-mediated
transport, in particular Table IV, and a list of STA-mediated transport experiments in Table V of \cite{Guery2019}.

Shortcuts provide  ideal trajectories for the control parameters but the results may be affected by noise and
implementation imperfections that limit experimentally the coherence of the transport, visibilities, and fidelities.
Ruschhaupt et al \cite{Ruschhaupt2012} introduced a ``noise sensitivity'' to quantify these effects as the second order term in the expansion of the final energy with respect to the perturbative noise, and demonstrated that the time dependence of the controls may be optimized
to achieve robust protocols in operations on two-level systems, see also
\cite{Lu2013,Daems2013}. Lu and coworkers \cite{Lu2014,Lu2018}  studied the effect of spring-constant noise  on STA-driven  transport of trapped ions, distinguishing  two types of contributions to the sensitivity:  static (independent of trap motion) and dynamical, with opposite behavior with respect to shuttling time. They also  demonstrated that the excitation can be reduced by proper process timing and design of the trap trajectory.

In this work we shall find the sensitivities for STA-mediated transport of atoms in optical lattices
with respect to noises in the three parameters of a
moving optical lattice potential  $A\sin^2(K x +\Phi)$, namely, noises  in the ``amplitude'' $A$,
in the phase  $\Phi$,
or in the wavenumber $K$, which affect, respectively, the trap depth, the trap position, and the lattice periodicity. Interestingly they have different effects and behavior, in particular with respect to static and dynamical components. This information will be instrumental in identifying
dominant sources of noise and to mitigate their effects.  To focus on the effect of these noises excluding other phenomena
and to get analytical results with explicit dependences, we shall assume throughout the paper conditions such that a  single atom is trapped in a given lattice site minimum, with negligible tunneling, interatomic interactions, and spontaneous emission. 
Internal-state dependence of the lattice is disregarded, in fact the internal state plays no role
in the following and it is assumed to remain unchanged along the  shuttling.
Moreover a deep lattice is assumed, in a Lamb-Dicke regime where the relevant atomic motion is effectively governed by a harmonic trap.
This last condition could  be  relaxed as explained in the final discussion.

In Sec. \ref{II} we review for completeness the invariant-based inverse engineering of STA trap trajectories  for a harmonic trap and
the general form of the noise sensitivities for a transport protocol.
In Sec. \ref{IV}, we  consider the three types of noise
for $A$, $K$, and $\Phi$. The noise spectrum may be arbitrary, but we pay special attention
to the white noise limit and to Ornstein-Uhlenbeck noise as a simple generalization to account for the effect of 
colored noise with a non-zero correlation time.
%
 
\section{Invariant-based inverse engineering and noise sensitivities\label{II}}
\subsection{Invariant-based inverse engineering \label{invasec}}
Let us first review the basic dynamical equations for a particle of mass $m$ trapped in 
a harmonic trap with angular frequency $\Omega(t)$ whose center moves along 
an arbitrary  trajectory $Q(t)$. Then we shall use this information to inverse engineer a special  trajectories  $q_0(t)$   that shuttle the particle without final excitation \cite{Torrontegui2011}.
Effective one-dimensional configurations are assumed throughout.
The Hamiltonian  in coordinate ($x$) representation  is
\beq
{\cal H}_0 (t)=\frac{{p}^2}{2 m} + \frac{1}{2}m \Omega^2(t)
[{x}-Q(t)]^2,
\label{h0}
\eeq
where ${p}$ is the   momentum 
operator.  We may subtract the purely time-dependent term and use instead
${H}_0={\cal H}_0 (t)-m\Omega^2(t)Q(t)^2/2$ to find the dynamics,
\beq\label{vqt}
{H}_0(t)=\frac{{p}^2}{2 m}-F(t){x}+\frac{m}{2}\Omega^2(t){x}^2. 
\eeq
$F(t)=m\Omega^2(t)Q(t)$ is a homogeneous force throughout space.

This Hamiltonian has a quadratic
Lewis-Riesenfeld invariant of the form \cite{Torrontegui2011,LR,LL,DL}
\beqa
\label{inva}
{I}(t)\!&=&\!\frac{1}{2m}\{\rho(t)[{p}-m\dot{q}_c(t)]-m\dot{\rho}(t)[{x}-\dot{q}_c(t)]\}^2
\nonumber\\
&+&\frac{1}{2}m \omega_0^2 \bigg[\frac{{x}-q_c(t)}{\rho(t)}\bigg]^2,
\eeqa
where $\omega_0$ is a constant, and ``invariance'' means that its expectation values remain constant for the states driven by ${H}_0$,  i.e.,
\beq
\frac{d {I}(t)}{d t} \equiv \frac{\partial {I}(t)}{ \partial t} +\frac{1}{i \hbar} [{I}(t), {H}_0(t)] =0.
\label{invadef}
\eeq
Assuming a quadratic-in-momentum ansatz for ${I}(t)$ in this equation, it is found that  $\rho(t)$ and $F(t)$ must satisfy the ``Ermakov'' and ``Newton''  equations
\beqa
\ddot{\rho}(t)+\Omega^2(t)\rho&=&\frac{\omega_0^2}{\rho^3(t)},
\nonumber\\
\ddot{q}_c(t)+\Omega^2(t)q_c(t)&=&F(t)/m.
\label{newton eq}
\label{erma}
\eeqa
Hereafter we conveniently choose $\omega_0=\Omega(0)$. 
$\rho(t)$ is a scaling factor that determines the width of the eigenstates of the invariant
and $q_c(t)$ is a classical trajectory for the forced oscillator, see Eq. (\ref{newton eq}).  The eigenstates of ${I}(t)$, Eq. (\ref{inva}),  
are centered at $q_c(t)$.
The eigenvalues $\lambda_n$ of ${I}(t)$ are  constant,
${I}(t)\psi_n(t)=\lambda_n \psi_n(t)$, whereas the (orthogonal) eigenstates of the invariant, $\psi_n(t)$,
are time dependent,
\beq\label{psinqt}
\psi_n(x,t)=\frac{1}{\sqrt{\rho}}e^{\frac{im}{\hbar}[\frac{\dot\rho x^2}{2\rho}
+\frac{(\dot{q}_c\rho-\dot{\rho}q_c)x}{\rho}]}\phi_n\bigg(\frac{x-q_c}{\rho}\bigg),
\eeq
where $\phi_n(x)$ are the eigenstates of a static harmonic oscillator with angular frequency $\omega_0$.
Arbitrary solutions of the time-dependent Schr\"odinger equation $i \hbar
\partial_t\Psi (x,t) = {H}_0(t) \Psi (x,t)$ may be expanded using
the ``transport modes'' $\Psi_n(x,t)\equiv e^{i\theta_n(t)} \psi_n(x,t)$, where
the Lewis-Riesenfeld phases $\theta_n(t)$ are  found so that  each transport mode is
itself a solution,
\beq
\label{LRphase}
\theta_n(t) = \frac{1}{\hbar} \int_0^t \Big\langle \psi_n (t') \Big|
i \hbar \frac{\partial }{ \partial t'} - {H}_0(t') \Big| \psi_n (t')  \Big\rangle d t'.
\eeq
Thus,
$
\Psi(x,t) = \sum_n c(n)e^{i\theta_n(t)} \psi_n(x,t),
$
where the $c(n)$ are time independent, and $n=0,1,...$.

In a rigid harmonic trap we may simply set
\beq
\Omega(t)=\omega_0,~~~\rho(t)=1.
\eeq
To inverse engineer
a trap trajectory $q_0 (t)$ that would  transport the particle without final excitations
from   $q_0(0) =0$ to $q_0 (T) =d$ in a time $T$, we shall design first
$q_c (t)$ and deduce $q_0(t)$ from the Newton equation (\ref{newton eq}) 
with $F(t)=m\omega_0^2 q_0(t)$.  We impose 
the boundary conditions \cite{Torrontegui2011}
\beqa
\label{conq}
q_0(0)=q_c(0)=0,~~ \dot{q}_c(0)=0,
\nonumber
\\
 q_0(T)=q_c(T)=d,~~\dot{q}_c(T)=0,
\eeqa
so that  
${I}(t)$ and
${H}_0(t)$ commute at $t=0$ and $t=T$. Therefore the two operators  share eigenvectors
at those times 
and the initial eigenvectors evolve into  final eigenvectors. Moreover,
the continuity of  $q_0(t)$ is guaranteed by the additional conditions
\beq \label{conqdd}
\ddot{q}_c(0)=0,~~\ddot{q}_c(T)=0.
\eeq
Note that the freedom to interpolate $q_c(t)$ in different ways between the trajectory boundaries can be used to 
produce  different shortcuts.   
\subsection{Noise sensitivity \label{III}}
Here we shall define noise sensitivities following \cite{Lu2018} but for a more general scenario, namely,
for a Hamiltonian  (\ref{h0}) where both $\Omega(t)$ and $Q(t)$ could be affected by  classical noise.
The origin of the noise in the harmonic model is that, as explained in the next section in detail,  different  parameters of the optical lattice potential may suffer from some noisy deviation from the ideal value. 
This deviation is represented by  $\lambda \xi(t)$, possibly multiplied by some appropriate dimensional factor depending on the parameter. 
$\lambda$ is the dimensionless perturbative parameter that should be set to one at the end of the calculation, and 
$\xi(t)$
is also dimensionless.
$\xi(t)$ is assumed to be unbiased, i.e., the  average over noise realizations  $\cE[\cdots]$ gives zero, 
and the (dimensionless) correlation function $\alpha$ is stationary,
\beq\label{xi}
\cE[\xi(t)]=0,~~~\cE[\xi(t)\xi(s)]=\alpha(t-s).
\eeq
We  also assume that there is no noise at initial time, so the initial conditions
for $\rho(t)$ and $q_c(t)$ are fixed as
\beqa\label{in condition}
\rho(0)&=&1, ~\dot{\rho}(0)=\ddot{\rho}(0)=0,
\nonumber\\
q_c(0)&=&0,~\dot{q}_c(0)=\ddot{q}_c(0)=0.
\eeqa
Now the auxiliary functions $\rho(t)$ and $q_c(t)$  are expanded in powers of $\lambda$,
\beqa\label{expansion}
\rho(t)&=&\rho^{(0)}(t)+\lambda \rho^{(1)}(t)+\cdot\cdot\cdot,
\nonumber\\
q_c(t)&=&q_c^{(0)}(t)+\lambda q_c^{(1)}(t)+\cdot\cdot\cdot.
\eeqa
Assuming as well a series expansion of $\Omega(t)$ and $Q(t)$ in $\lambda$,  we get in 
zeroth order  (noiseless limit)
\beqa\label{0 order}
\rho^{(0)}(t)&=&1,\nonumber\\
\ddot{q}_c^{(0)}(t)+\omega_0^2q_c^{(0)}(t)&=&\omega_0^2q_0(t),
\eeqa
where $q_c^{(0)}(t)$ satisfies  Eqs. (\ref{conq}) and (\ref{conqdd}).

We also assume that there is no noise at the final time, ${\cal{H}}(T)={p}^2/2m+ m\omega_0^2({x}-d)^2/2$.
The  expectation value of ${\cal{H}}(T)$ for a state $\Psi_n(T)=e^{i\theta_n(T)}\psi_n(T)$, see Eq. (\ref{psinqt}),
that started as the $n_{th}$ mode for a  realization of the noise $\xi(t)$ can be found exactly,
\beqa\label{energy}
E_{n,\xi}&=&\la {{\cal H}}(T)\ra=\la \Psi_n(T)|{\cal H}(T)|\Psi_n(T)\ra
\nonumber\\
&=&\frac{m}{2}\omega_0^2[q_c(T)-d]^2+\frac{\hbar\omega_0}{4}(2n+1)\frac{1+\rho^4(T)}{\rho^2(T)}
\nonumber\\
&+&\frac{m}{2}\dot{q}_c^2(T)+\frac{\hbar}{4\omega_0}(2n+1)\dot{\rho}^2(T).
\eeqa
$E_{n,\xi}$ can be expanded  in powers of $\lambda$ as
\beq
E_{n,\xi}\approx E_{n,\xi}^{(0)}+\lambda E_{n,\xi}^{(1)}+\lambda^2 E_{n,\xi}^{(2)}+\cdot\cdot\cdot,
\eeq
with  $E_{n,\xi}^{(1)}=\frac{\partial E_{n,\xi}}{\partial\lambda}$, $E_{n,\xi}^{(2)}=\frac{1}{2}\frac{\partial^2 E_{n,\xi}}{\partial\lambda^2}$.
Combining Eq. (\ref{energy}) and the expansions  for $\rho(t)$ and $q_c(t)$ in Eq. (\ref{expansion}), we find
the zeroth order $E_{n,\xi}^{(0)}=\hbar\omega_0(n+\frac{1}{2})$ and  $E_{n,\xi}^{(1)}=0$, as expected, as well as
\beqa
E_{n,\xi}^{(2)}&=&\frac{1}{2}m\omega_0^2q_c^{(1)}(T)^2+\hbar\omega_0(2n+1)\rho^{(1)}(T)^2
\nonumber\\
&+&\frac{1}{2}m\dot{q}_c^{(1)}(T)^2+\frac{\hbar\dot{\rho}^{(1)}(T)^2}{4\omega_0}(2n+1).
\eeqa
Averaging over different realizations of the noise,
\beq\label{En}
E_n=\cE[ E_{n,\xi}]= E_n^{(0)}+\lambda^2 \frac{1}{2}\cE\bigg[\frac{\partial^2 E_{n,\xi}}{\partial\lambda^2}\bigg],
\eeq
where $E_n^{(0)}=E_{n,\xi}^{(0)}$.

The noise sensitivity for a given transport protocol is defined as the second order coefficient, so it has dimensions of energy,
\beqa
G(T;n) &=& \frac{1}{2}\cE\bigg[\frac{\partial^2 E_{n,\xi}}{\partial\lambda^2}\bigg]=\cE[E_{n,\xi}^{(2)}]
\nonumber\\
&=&G_1+G_2.
\eeqa
We have separated the contributions related to $\rho$ and to $q_c$,
\beqa\label{sensitivity}
G_1&=&\hbar(2n+1)\left\{\omega_0\cE[\rho^{(1)}(T)^2]
+\frac{1}{4\omega_0}\cE[\dot{\rho}^{(1)}(T)^2]\right\},
\nonumber\\
G_2&=&\frac{1}{2}m\omega_0^2\cE[q_c^{(1)}(T)^2]+
\frac{1}{2}m\cE[\dot{q}_c^{(1)}(T)^2].
\eeqa
In the following, we will discuss three different kinds of noise in the moving optical lattice and find the exact expressions of the corresponding sensitivities.

\section{Noise in a moving optical lattice \label{IV}}
Let us consider an effective  potential of the form
\beqa
\label{sw}
V=A\sin^2[K x +\Phi(t)]
\eeqa
due to a laser standing wave.
All three coefficients could be affected by noise around central values $a$, $k$ and $\phi$ so it is useful to consider an auxiliary 
``noiseless version'' of Eq. (\ref{sw}),
\beqa
\label{swnf}
V({\rm noise\,free})=a\sin^2[k x +\phi(t)].
\eeqa
Among the periodic minima we pick up the one at  $Q(t)=-\Phi(t)/K$ as the one ``occupied'' by an atom.
Expanding around this point we find a quadratic
approximation for Eq. (\ref{sw}),
\beqa
A\sin^2[Kx+\Phi(t)]&\approx&AK^2[x-Q(t)]^2,
\eeqa
where $A$ is the potential depth of the lattice and $K$ is the wavenumber of the laser light.
Considering the possible time dependences,  noisy or otherwise, the quadratic Hamiltonian may be  written as
Eq. (\ref{h0}) with  $\frac{1}{2}m\Omega(t)^2=AK^2$. 
Without any noise  $\Omega(t)=\omega_0$,   $\frac{1}{2}m\omega_0^2=ak^2$, and $Q=q_0$. 

\subsection{Wavenumber (accordion)  noise}
%
%
%
\begin{figure}[t]
\begin{center}
\scalebox{0.7}[0.7]{\includegraphics{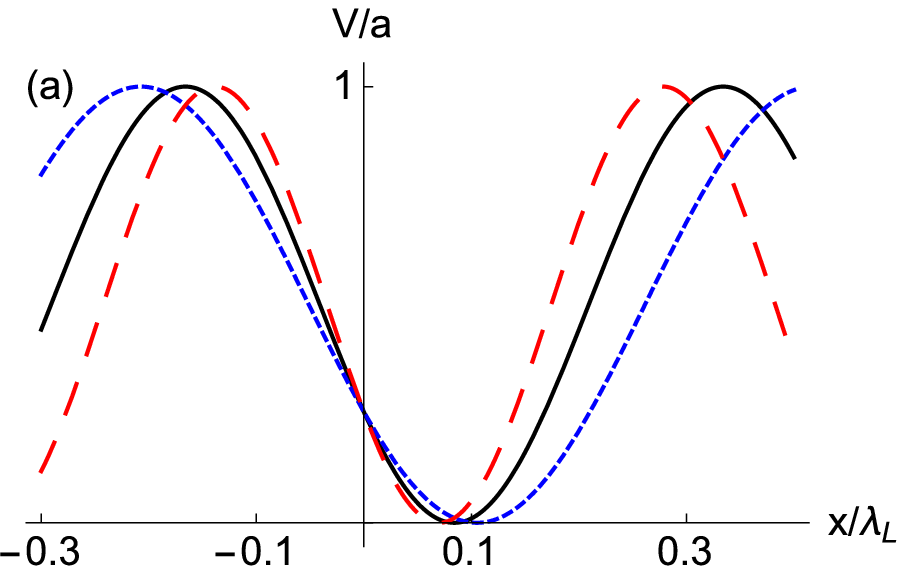}}
\scalebox{0.7}[0.7]{\includegraphics{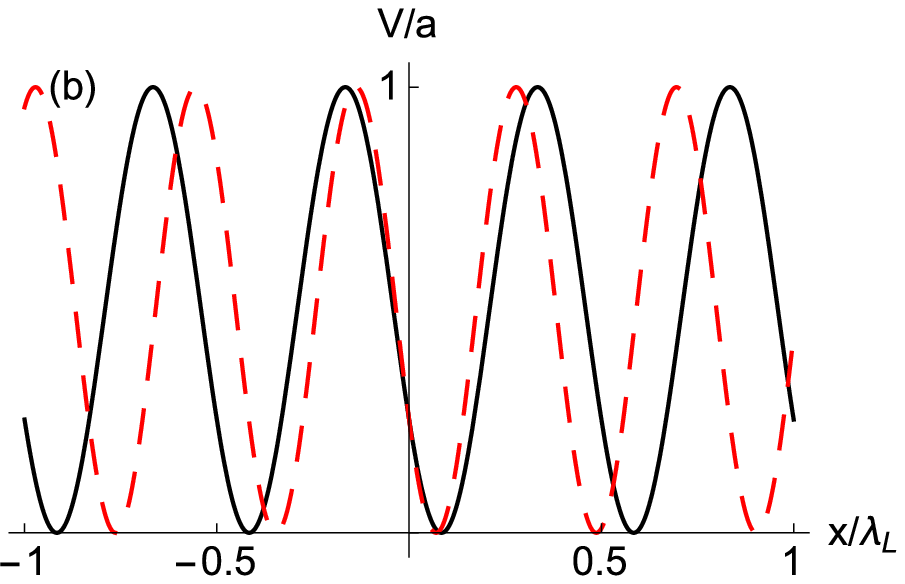}}
\caption{(Color
online) Schematic effect of accordion ($K$)  noise. Accordion noise consists of random compressions/expansions
with respect to the pivot point $x=0$.  (a) In a particular  minimum, the one at $q_0>0$ without noise,  expansions imply smaller trap frequencies together with  displacements to the right, and compressions the opposite phenomena. The displacements of the minimum due to $K$ noise
increase with the distance to the pivot.
The black solid line is the noiseless trap at some time during transport. The red dashed line represents a compression and the blue dotted line an expansion.
The parameter values are chosen to easily visualize the effect and do not intend to be realistic.
(b) Several lattice periods for the reference potencial without noise (black solid line) and the compressed version (red dashed line).}\label{figk1}
\end{center}
\end{figure}
Accordion lattices have been implemented in different ways \cite{Li2008,Williams2008,AlAssam2010,Tao2018}
to change the lattice periodicity  keeping other parameters fixed.
We consider first that the wave vector suffers from an involuntary ``accordion noise'' as $K=k[1+\lambda \xi(t)]$, whereas 
$A=a$ and $\Phi=\phi$. 
Some possible realizations of the potential at a given time are depicted in Fig. \ref{figk1} for a particular well (a) or for several 
wells (b). The harmonic potential with $K$ noise now can be written as
\beqa
V&=&\!\!\!ak^2\bigg[1+\lambda \xi(t)\bigg]^2\bigg[x+\frac{1}{1+\lambda\xi(t)}\frac{\phi(t)}{k}\bigg]^2\nonumber\\
&=&\!\!\!\frac{1}{2}m\omega_0^2\bigg[1+\lambda \xi(t)\bigg]^2\bigg[x\!-\frac{q_0(t)}{1+\lambda \xi(t)}\bigg]^2\nonumber\\
&=&\frac{1}{2}m\Omega^2(t)[x-Q(t)]^2,
\eeqa
where $\Omega^2(t)=\omega^2_0[1+\lambda\xi(t)]^2$, whereas the minimum at $Q(t)={q_0(t)}/({1+\lambda\xi(t)})$ is displaced
by the noise proportionally to $q_0(t)$.
Both the spring constant and the trap position  are
affected  by the accordion noise.

Substituting the expansions of $\rho(t)$ and $q_c(t)$ of Eq. (\ref{expansion}) into Eq. (\ref{erma}),
and keeping only the first order in $\lambda$, $\rho^{(1)}(t)$ and $q_c^{(1)}(t)$ will satisfy
\beqa\label{rho-qc-k}
\ddot{\rho}^{(1)}(t)+4\omega_0^2\rho^{(1)}(t)\!\!&=&\!\!-2\omega_0^2\xi(t),
\nonumber\\
\ddot{q}_c^{(1)}(t)+\omega_0^2q_c^{(1)}(t)\!\!&=&\!\![\ddot{q}_{c}^{(0)}(t)-\omega_0^2q_{c}^{(0)}(t)]\xi(t),
\eeqa
with initial conditions $\rho^{(1)}(0)=\dot{\rho}^{(1)}(0)=\ddot{\rho}^{(1)}(0) $ and $q_c^{(1)}(0)=\dot{q}_c^{(1)}(0)=\ddot{q}_c^{(1)}(0)$.
  
The solutions of Eq. (\ref{rho-qc-k}) are
\beqa\label{k solution}
\rho^{(1)}(t)&=&-\omega_0\int_0^tds\, \xi(s)\sin[2\omega_0(t-s)],
\nonumber\\
q_c^{(1)}(t)&=&\frac{1}{\omega_0}\int_0^tds\, \xi(s)[\ddot{q}_{c}^{(0)}(s)-\omega_0^2q_{c}^{(0)}(s)]\sin[\omega_0(t-s)].
\nonumber\\
\eeqa
Substituting them  into Eq. (\ref{sensitivity}), we get the sensitivity
\beqa
G(T;n)\!\!&=&\!\!G_{1K}(T;n)+G_{2K}(T;n),\nonumber\\
G_{1K}(T)\!\!&=&\!\!\hbar\omega_0^3(4n+2)\int_0^Tds\ \alpha(s)(T-s)\cos(2\omega_0s),
\nonumber\\
G_{2K}(T)\!\!&=&\!\!m\!\int_0^T\!\!\!ds\ \alpha(s)f_K(s),
\label{G for k}
\eeqa
where
\begin{equation}
f_K(s)=\cos(\omega_0s)\int_0^{T-s}B(u)B(u+s)du,
\label{fsK}
\end{equation}
with $B(u)= \ddot{q}_{c}^{(0)}(u)-\omega_0^2q_{c}^{(0)}(u)$.
$G_{1K}$ is independent of the trajectory, it is a ``static'' contribution  that  depends on $n$, the frequency $\omega_0$, the correlation 
function of the noise $\alpha(t)$, and shuttling time $T$.
Instead, $G_{2K}$ is a ``dynamical'' contribution that depends on the trajectory, on $\alpha(t)$,  and on the mass $m$.
The static/dynamical character can be traced back to Eq. (\ref{rho-qc-k}). The noise forcing term in the equation for $\rho^{(1)}$ does not depend on the trajectory whereas the one for $q_c^{(1)}$ does. 
However $G_1$ and $G_2$ in Eq. (\ref{sensitivity}) do not necessarily become, respectively,  static and dynamical sensitivities 
for all noises as they do here, see in particular  Sec. \ref{Qnoise} on ``position noise'' below.  Each noise type requires a separate analysis.

\begin{figure}[t]
\begin{center}
\scalebox{0.84}[0.84]{\includegraphics{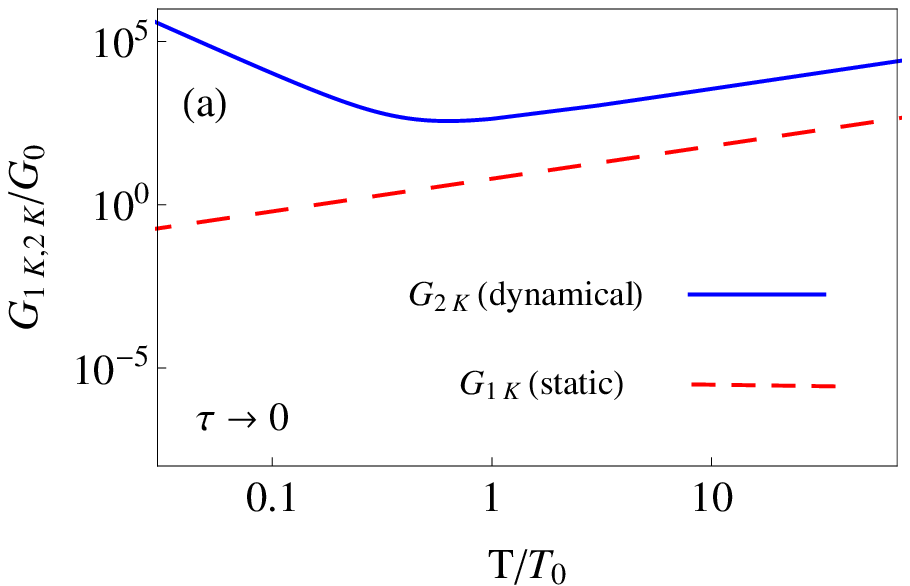}}
\scalebox{0.84}[0.84]{\includegraphics{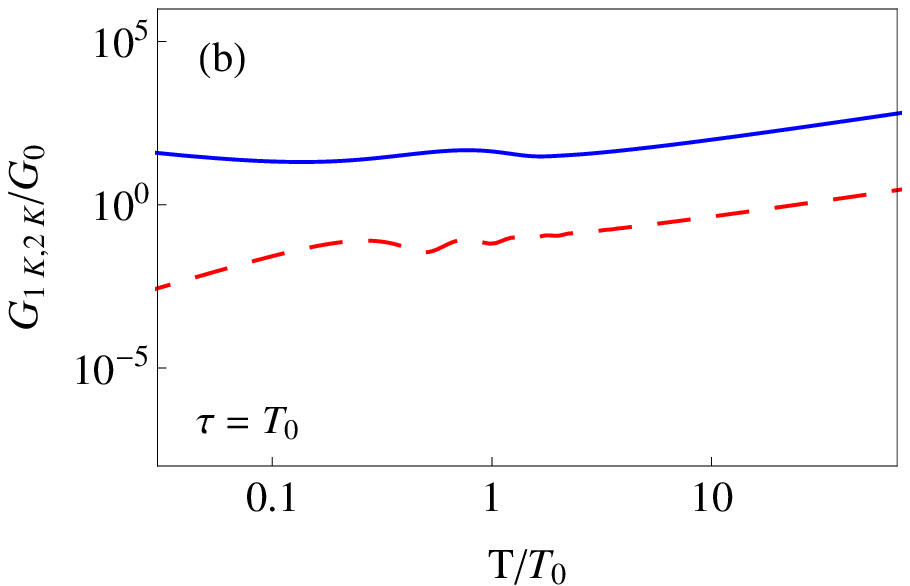}}
\scalebox{0.84}[0.84]{\includegraphics{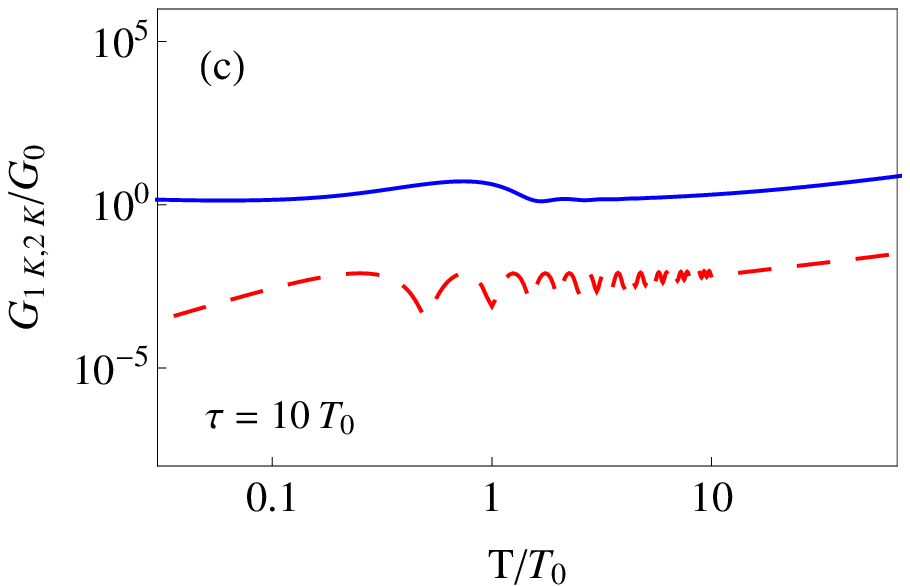}}
\caption{(Color
online) Log-log plot of the sensitivity to accordion noise in units  of $G_0=\hbar\omega_0^2D$ for a polynomial protocol versus final time in units of the oscillation period
$T_0=2\pi/\omega_0$ and for different values of $\tau$: (a) $\tau\to 0$ (white noise limit); (b) $\tau=T_0$; (c) $\tau=10 T_0$.
The blue solid line is the dynamical component $G_{2K}$ and the red dashed line the static component $G_{1K}$. 
The parameters are $\lambda_{L}=2\pi/k=866$ nm, $d=\frac{1}{2}\lambda_L$, $a=850\, E_R$,
mass of ${}^{133}$Cs, $n=0$,
$\omega_0=\sqrt{2a k^2/m}=2\pi\times 116$ kHz, and recoil energy $E_R=(\hbar k)^2/(2m)$.
The same scale is kept in these  three figures and in later figures for the other noises (Figs. \ref{figa3} and \ref{figphi2})  to
compare easily the different sensitivities.  
}
\label{figk3}
\end{center}
\end{figure}

To evaluate the integrals in Eq. (\ref{G for k}) the correlation function $\alpha(t)$ of the noise  has to be specified. 
We consider Ornstein-Uhlenbeck (OU) noise with correlation function
\beq
\alpha(t)=\frac{D}{2\tau}e^{-t/\tau}
\eeq
as a simple, natural generalization
of Gaussian white noise  to introduce a finite correlation time $\tau$.
$D$, with dimensions of time, sets the strength of the noise (the factor $D$ was taken out of the correlation function in \cite{Lu2018}.\footnote{When comparing  
the present work and \cite{Lu2018} note also that $\lambda$ had dimensions of square root of time there, whereas it is dimensionless here.}  The convention here is as in \cite{Lu2014}.)
OU noise is not the most general colored noise, but it
covers a much larger domain  than the white-noise assumption
\cite{Lehle2018}.
When $\tau\rightarrow 0$, it reduces to white noise, and is also
instrumental in generating flicker noise by superposing a range of correlation times \cite{Lu2014}.

To be more specific and see the behavior of the sensitivity, we  assume  a simple polynomial ansatz, $q_c^{(0)}(t)=\sum_{j=0}^{5}b_jt^j$, where the $b_j$ are fixed to satisfy the imposed boundary conditions. The optical lattice  moves in our simulations from $0$ to $d=\lambda_L/2$, where $\lambda_L$ is the wavelength of the light creating the optical lattice, so that $d$
is the distance between two contiguous minima. In
Fig. \ref{figk3}, the sensitivity components $G_{1K}$ and $G_{2K}$ are shown versus final time for a Cs atom, see further details in the caption. 
The lattice parameters are realistic and taken from \cite{Belmechri2013}. They correspond to a Lamb-Dicke regime, $\hbar\omega_0/E_R\approx 58$, where $E_R=(\hbar k)^2/(2m)$ is the recoil energy.   

In Fig. \ref{figk3}  we include small $T$ values below the period $T_0=2\pi/\omega_0$ for completeness, but note that the harmonic and single well approximations will  fail in such a regime.
For a  simple estimate of minimal allowed shuttling times we may compare a lower bound for averaged potential energy during transport \cite{Torrontegui2011},
with the potential depth $a$, i.e., 
$6 m d^2/(T^4 \omega_0^2a)\gg 1$ should hold for the particle to stay in a minimum.  Using $\omega_0=\sqrt{2a k^2/m}$ and $d=\pi/k$ gives a minimal time scale $T\approx  T_0/2$.
Shorter times which would not be affected by the failure of the harmonic approximation may be implemented by applying a
time-dependent homogeneous force compensating  the inertial force, this is discussed briefly in the final section, see also \cite{Torrontegui2011}.

In the white noise limit $\tau\to 0$  
Eq. (\ref{G for k}) gives
\beqa\label{white limit for k}
G_{1K}&=&\hbar\omega_0^3 D {(2n+1)T},
\nonumber\\
G_{2K}&=&m d^2D\Bigg(\frac{181}{924}\omega_0^4T+\frac{60}{7T^3}+\frac{10\omega_0^2}{7T}\Bigg),
\eeqa
which implies a minimum for the dynamical term $G_{2K}$ at $T\approx 0.63$ $T_0$ and a monotonous growth with process time $T$  
for the static part $G_{1K}$. For $T>T_0$ both terms grow linearly with time $T$ as shown in the right part of  Fig. \ref{figk3} (a). 
Comparing $G_{1K}$ and the linear part of $G_{2K}$ we find that for this noise $G_{2K}$ is always dominant 
in the Lamb-Dicke regime. In the white noise limit, with $d=\pi/k$,    
\beq
\frac{G_{2K}({\rm linear\, in\,} T\, {\rm term})}{G_{1K}}=\frac{181}{924}\frac{m \omega_0 d^2}{\hbar}\approx 0.96 \frac{\hbar\omega_0}{E_R}.
\eeq
The effect of a finite correlation time with a OU correlation function is explored numerically in Figs. \ref{figk3} (b) and (c): increasing correlation times diminish the sensitivity 
in all time $T$ regions and  even suppress strongly the short-time $T$ growth of sensitivity characteristic of the white noise limit. 
$G_{2K}$ stays dominant over $G_{1K}$ for all $\tau$.   

The above results are consistent with known effects of spring-constant  noise  in static traps \cite{Savard1997,Gehm1998}. 
For the static part alone, i.e., assuming no transport, $q_0(t)=0$, and for $T\gg\tau$,
\beq
\frac{d E_n}{dT}=4 \omega_0^2 \pi E_n^{(0)}S_K(2\omega_0),
\eeq
where $S_K(2\omega_0)$ is the spectral density  for the fractional  fluctuation in the wavenumber
at the second harmonic of the trap (we have set $\lambda=1$),
\begin{equation}
S_K(2\omega_0) = \frac{1}{\pi} \int_0^\infty  \alpha(t) \cos(2\omega_0 t)  dt,
\label{Sk}
\end{equation}
see also the corresponding discussion for amplitude noise in the following subsection.

\subsection{Amplitude (trap depth) noise\label{anoise}}
%
%
%
%
\begin{figure}[t]
\begin{center}
\scalebox{0.8}[0.8]{\includegraphics{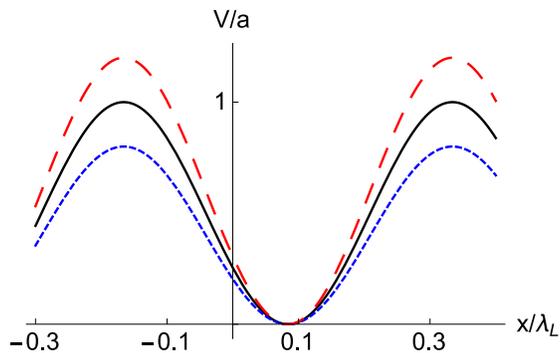}}
\caption{(Color
online) Two realizations of the potential due to amplitude noise (red dashed line and dotted blue line) at some given time. The corresponding noiseless potential is also represented as a solid black line.}\label{figa1}
\end{center}
\end{figure}
Trap depth noise may be due to laser intensity fluctuations as well as to pointing instabilities of the laser beams that could arise
as a consequence of shifts of the laser beam, acoustic vibrations or air flow \cite{Kuhr2003}. For example Kuhr et al. \cite{Kuhr2003}, estimated
the fluctuations of the trap depth in their optical lattice setting to reach up to $3\%$ for time scales $t>100$ ms.
We consider  amplitude noise as $A=a[1+\lambda \xi(t)]$ (whereas $K=k$, and $\Phi=\phi$), see Fig. \ref{figa1}, so that the optical lattice potential can be written as
\beqa
V=a[1+\lambda  \xi(t)]k^2(x-q_0)^2=
\frac{1}{2}m\Omega^2(t)(x-q_0)^2,
\eeqa
where $\Omega^2(t)=\omega_0^2[1+\lambda \xi(t)]$ is affected by a classical spring constant noise.
%

\begin{figure}[t]
\begin{center}
\scalebox{0.84}[0.84]{\includegraphics{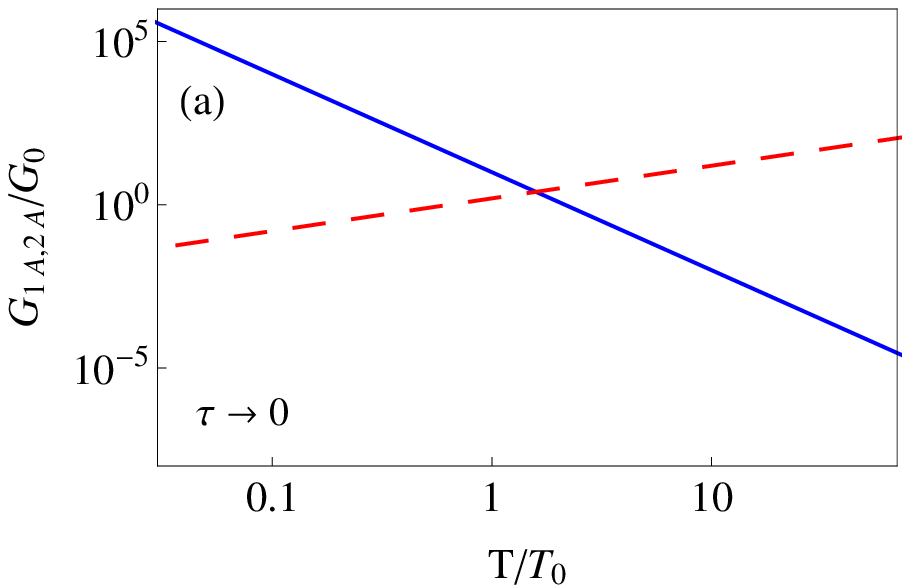}}
\scalebox{0.84}[0.84]{\includegraphics{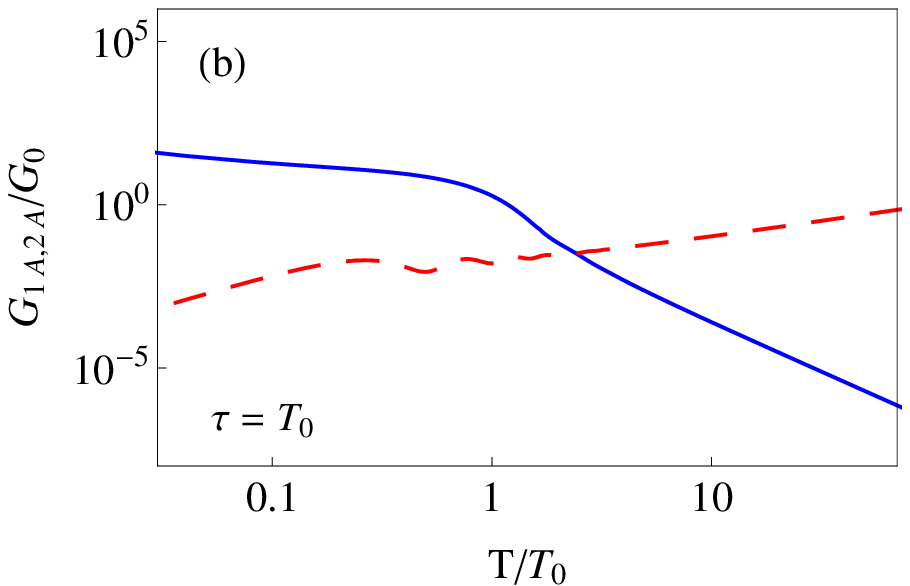}}
\scalebox{0.84}[0.84]{\includegraphics{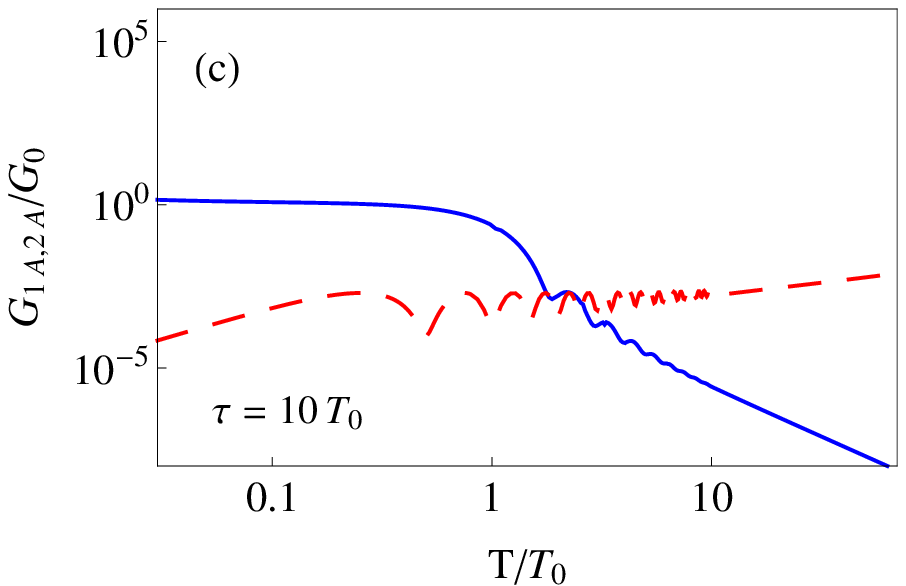}} 
\caption{(Color
online) Amplitude-noise sensitivity for a polynominal protocol versus final time (log-log plot) and for different correlation times $\tau$.
Dashed red line: static term $G_{1A}$; solid blue line: dynamical term $G_{2A}$. 
The parameters and scales are the same as in Fig. \ref{figk3}. 
}\label{figa3}
\end{center}
\end{figure}
Similarly to the procedure followed for accordion noise,  substituting  the expansions of $\rho(t)$ and $q_c(t)$
into Eq. (\ref{erma}),
and keeping only the first order of $\lambda$, $\rho^{(1)}(t)$ and $q_c^{(1)}(t)$ will satisfy
\beqa\label{rho-qc-A}
\ddot{\rho}^{(1)}(t)+4\omega_0^2\rho^{(1)}(t)&=&-\omega_0^2\xi(t),
\nonumber\\
\ddot{q}_c^{(1)}(t)+\omega_0^2q_c^{(1)}(t)&=&\ddot{q}_c^{(0)}(t)\xi(t),
\eeqa
with initial conditions $\rho^{(1)}(0)=\dot{\rho}^{(1)}(0)=\ddot{\rho}^{(1)}(0) $ and $q_c^{(1)}(0)=\dot{q}_c^{(1)}(0)=\ddot{q}_c^{(1)}(0) $.
The solutions of Eq. (\ref{rho-qc-A}) are
\beqa\label{spring solution}
\rho^{(1)}(t)&=&-\frac{\omega_0}{2}\int_0^tds\,\xi(s)\sin[2\omega_0(t-s)],
\nonumber\\
q_c^{(1)}(t)&=&\frac{1}{\omega_0}\int_0^tds\, \xi(s)\sin[\omega_0(t-s)]\ddot{q}_c^{(0)}(s).
\eeqa

Substituting  $\rho^{(1)}(t)$ and $q_c^{(1)}(t)$  into Eq. (\ref{sensitivity}), we get
\beqa
G(T;n)&=&G_{1A}(T;n)+G_{2A}(T;n),\nonumber\\
G_{1A}(T)&=&{\hbar\omega_0^3}\bigg(n+\frac{1}{2}\bigg)\!\!\int_0^T\!\!ds\ \alpha(s)(T\!-\!s)\cos(2\omega_0s),
\nonumber\\
G_{2A}(T)&=&{m}\!\int_0^T\!\!\!ds\ \alpha(s)f_A(s),
\label{G for ampli}
\eeqa
where
%
%
\begin{equation}
f_A(s)=\cos(\omega_0s)\!\int_0^{T-s}\!\!du\ \ddot{q}_c^{(0)}(u)\ddot{q}_c^{(0)}(u+s).
\label{fsA}
\end{equation}
As before we compute the integrals for OU noise, and use the polynomial ansatz for $q_c$. 
In the white noise limit $\tau\to 0$  
\beqa
G_{1A}&=&\frac{D}{4}\hbar\omega_0^3(2n+1)T,
\nonumber\\
G_{2A}&=&\frac{D60md^2}{7T^3}.
\label{g2a}
\eeqa
Up to the scaling due to the optical lattice parameters, these expressions coincide with the results given in  \cite{Lu2018}
for ``spring-constant noise'', and different limits and regimes were discussed there in detail.
Here we note that different from the accordion noise sensitivities, $G_{1A}$ (static) and $G_{2A}$ (dynamical) behave in  opposite ways to each other in all $T$ domains, 
and cross at a special optimal time with minimal sensitivity, see Fig. \ref{figa3}.

The static part alone (no transport, $q_0(t)=0$) implies for $T$ larger than the correlation time
a heating rate  in agreement with  \cite{Savard1997,Gehm1998},
\beq
\frac{d E_n}{dT}=\omega_0^2 \pi E_n^{(0)}S(2\omega_0),
\eeq
where $S_A(2\omega_0)$ is the spectral density  for the fractional  fluctuation in the amplitude (trap depth)
at the second harmonic of the trap,
\begin{equation}
S_A(2\omega_0) = \frac{1}{\pi} \int_0^\infty  \alpha(t) \cos(2\omega_0 t)  dt.
\label{SA}
\end{equation}
Eqs. (\ref{Sk}) and (\ref{SA}) are in fact equivalent since both $S_A$ and $4S_K$ may be interpreted as the spectrum for the
fractional fluctuation of the spring constant. 

The effect of increasing  $\tau$ using OU noise is again to diminish the sensitivities, and to suppress the growth of the dynamical 
sensitivity for  small $T<T_0$, see Fig. \ref{figa3}.  

\subsection{Phase (trap position) noise\label{Qnoise}}
%
%
%
%
\begin{figure}[t]
\begin{center}
\scalebox{0.8}[0.8]{\includegraphics{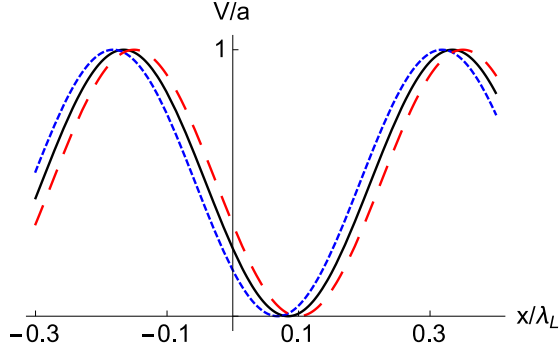}}
 \caption{(Color
online) Schematic representation of position noise in the optical lattice. The black solid line is the noiseless potential at some instant during the driving transport. The red dashed line and blue dotted line are two possible realizations of the potential due to position noise.}\label{figphi1}
\end{center}
\end{figure}
The standing wave phase $\phi(t)$ can be changed in time, moving the interference pattern, in different ways, see e.g. \cite{Schrader2001,Zemanek2019}:
one of the laser beams can be moved by mechanically moving a mirror \cite{Middelmann2012};  the phase of one of the laser beams can be controlled
with an electro-optical modulator; or a frequency mismatch $\Delta \nu$ between the counterpropagating beams controlled by acousto-optical modulators produces
a phase $\pi\Delta\nu t$. Of course all these methods are amenable to an imperfect control and fluctuations.
Here we consider  phase noise as $\Phi(t)=\phi(t)-\lambda\xi(t)$ independent of other possible noises ($A=a$, $K=k$), see Fig. \ref{figphi1}. The harmonic potential
takes now the form
\beqa
\!\!V\!=\!ak^2\!\bigg[x\!+\!\frac{\phi(t)\!-\!\lambda\xi(t)}{k}\bigg]^{\!2}
\!\!\!=\!\frac{m\omega_0^2}{2}\bigg[x\!-\!q_0(t)\!-\!\frac{\lambda}{k}\xi(t)\bigg]^{\!2}\!\!.
\eeqa
The phase noise implies noise in the trap position,  $Q(t)=q_0(t)+\frac{\lambda}{k}\xi(t)$.

First order equations are now 
\beqa\label{rho-qc-phi}
\ddot{\rho}^{(1)}(t)+4\omega_0^2\rho^{(1)}(t)&=&0,
\nonumber\\
\ddot{q}_c^{(1)}(t)+\omega_0^2q_c^{(1)}(t)&=&\frac{\omega_0^2}{k}\xi(t),
\eeqa
with initial conditions $\rho^{(1)}(0)=\dot{\rho}^{(1)}(0)=\ddot{\rho}^{(1)}(0) $ and $q_c^{(1)}(0)=\dot{q}_c^{(1)}(0)=\ddot{q}_c^{(1)}(0) $. 
The solutions of Eq. (\ref{rho-qc-phi}) are
\beqa\label{phi solution}
\rho^{(1)}(t)&=&0,
\nonumber\\
q_c^{(1)}(t)&=&\frac{\omega_0}{k}\int_0^tds\, \xi(s)\sin[\omega_0(t-s)],
\eeqa
\begin{figure}[t]
\begin{center}
\scalebox{0.9}[0.9]{\includegraphics{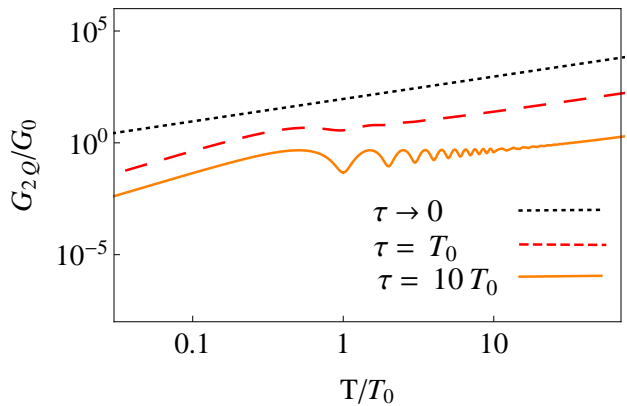}}
\caption{(Color
online) Position-noise sensitivity for a polynomial transport protocol versus final time (log-log plot).
The parameters and the scales are the same as in Fig. \ref{figk3}.
 }\label{figphi2}
\end{center}
\end{figure}
%
which give the sensitivities
\beqa
G(T;n)&=&G_{1Q}(T;n)+G_{2Q}(T;n),\nonumber\\
G_{1Q}(T)&=&0,
\nonumber\\
G_{2Q}(T)&=&\frac{m\omega_0^4}{k^2}\int_0^Tds\ \alpha(s)(T-s)\cos(\omega_0 s).
\label{G for phase}
\eeqa
The position noise sensitivity depends on the factor ${m\omega_0^4}/{k^2}$, $\alpha$, and $T$. 
There is only a static contribution which, for this noise, depends on $G_2$ rather than on $G_1$ as in the previous two noises.
Note also the independence on $n$ of $G_{2Q}$ unlike the static terms $G_{1K}$ and $G_{1A}$. 
For a transport process
the way to diminish its effect is to shorten the transport time.

As for the two previous noises we consider OU noise to compute the integral in Eq. (\ref{G for phase}). 
In the white noise limit, 
\beq
G_{2Q}=\frac{m  \omega_0^4}{2k^2}{DT}
\label{g2q}
\eeq
as shown in Fig. \ref{figphi2}. Increasing $\tau$ diminishes the sensitivity and also affects the slopes differently for $T$ larger or 
smaller than $T_0$.

For times $T$ larger than the correlation time we  find in second order, in agreement  with  \cite{Savard1997,Gehm1998},
the heating rate
\beq
\frac{d E_n}{dT}=m\omega_0^4 \pi S_{Q}(\omega_0),
\eeq
where $S_{Q}(\omega_0)$ is now the spectral density  for the fluctuation of the trap position (we set $\lambda=1$, otherwise multiply by $\lambda^2$),
\begin{equation}
S_{Q}(\omega_0) = \frac{1}{\pi}\int_0^\infty \frac{1}{k^2}\alpha(t) \cos(\omega_0 t)dt.
\end{equation}

\section{Discussion}
In this paper we have found the energy sensitivities with respect to  noise in a conveyor-belt optical lattice that moves according to shortcut-to-adiabaticity protocols to transport atoms. The three types of noise considered affect
the periodicity, the trap depth, or the trap position. A broad range of experimental settings  may lead to these noises, to different combinations,
or even to other noise forms (e.g. rocking). While the detailed analysis of the experimental settings is out of the scope of this work, the dependences found for the sensitivities will help to make a proper  diagnosis of the predominant noise type and to implement  mitigation strategies.
Position noise is only affected by the static sensitivity which grows linearly with the shuttling time independently of the trajectory so the noise effect can only be mitigated by
shortening the process time. Trap depth noise shows a more complex scenario for the sensitivity with a minimum
at a specific shuttling time  with dynamical effects dominating at very short times and static ones at long times. To locate the shuttling time
where the sensitivity is minimal the analysis in
\cite{Lu2018} for spring constant noise is applicable. Dynamical sensitivities can in principle be diminished  by optimizing
the trajectory, a task  left for future work.  Accordion noise is dominated by the dynamical sensitivity at all shuttling times which also shows
a minimum.

The existence of sensitivity minima demonstrates that the naive expectation that a smaller process time is always beneficial to combat the deleterious effects of noise is not necessarily true. Each type of noise requires a separate analysis and may or may not fulfill this expectation.
It is interesting to compare the dominant sensitivities due to different noises in the regime $T>T_0$. In all cases they grow linearly with time for white noise. In the Lamb Dicke regime the amplitude noise is found to have a weaker effect 
(although increased by $n$) than the other two, which behave similarly, see Eqs. (\ref{white limit for k},\ref{g2a},\ref{g2q}): 
${G_{2Q}}/{G_{1A}}={\hbar\omega_0}/[{E_R(2n+1)}]$, and 
${G_{2K}}\approx3.86\, G_{2Q}$.

A limitation of the shortcuts as implemented in Sec. \ref{invasec} is that  shuttling times shorter than an oscillation period break down the
simplifying conditions assumed (motion in a single harmonic well). Shorter-time shortcuts may however be applied by compensating
the inertial acceleration of the rigidly moving potential $U[q-q_c(t)]$ (the optical lattice potential) with an appropriate homogeneous force $-m\ddot{q}_c$ \cite{Masuda2010,Torrontegui2011}. This trick does not require the
trapping  potential $U$ to be harmonic, and  wavefunctions that are initially stationary stay so during the whole transport
in the  frame moving with $q_c$. In fact the effective potential in the moving frame stays stationary, ``nothing happens'' in that frame, apart from
possible noises.
Implementing this compensation may be  technically challenging and to the best of our knowledge it has not been implemented yet for
optical lattices, but the resulting benefits could make the effort worthwhile. We point out that lattice controlled rotations \cite{Williams2008,AlAssam2010} may be a way to implement the compensation.

Finally, the current noise analysis is also useful and applicable in the harmonic approximation
to other transport platforms and systems such as  atomic transport in
moving magnetic microtraps in chips \cite{Keil2016,Navez2016} or of ions in Paul traps \cite{Bowler2012,Walther2012,Kaufmann2018}.
\vspace*{.2cm}\\

\section*{Acknowledgement}
We are grateful to A. Alberti for helpful discussions. This work was supported by the Basque Country Government (Grant No. IT986-16), by   PGC2018-101355- B-I00 (MCIU/AEI/FEDER,UE), and  by the Key Research Project in
Universities of Henan Province (Grant No20B140016).

\label{Bibliography}
\bibliographystyle{unsrt}
\bibliography{Bibliography}
\end{document}